# Mid-Infrared Frequency Combs and Pulse Generation based on Single Section Interband Cascade Lasers

**Pavel Abajyan[1#], Baptiste Chomet[1], Daniel A. Diaz-Thomas[2], Mohammadreza Saemian[1], Martin Mičica[1], Juliette Mangeney[1], Jerome Tignon[1], Alexei N. Baranov[2], Konstantinos Pantzas[3], Isabelle Sagnes[3], Carlo Sirtori[1], Laurent Cerutti[2], Sukhdeep Dhillon[1*]**

[1] Laboratoire de Physique de l'Ecole normale supérieure, ENS, Université PSL, CNRS, Sorbonne Université, Université de Paris, 24 rue Lhomond, 75005 Paris, France

[2] IES, Univ. Montpellier, CNRS, F-34000 Montpellier, France

[3] C2N, CNRS-Univ. Paris-Sud, Univ. Paris-Saclay, 10 Avenue Thomas Gobert, F-91120 Palaiseau, France

# Corresponding author 1: pavel.abajyan@phys.ens.fr

*Corresponding author 2: sukhdeep.dhillon@ens.fr

**Abstract: Interband Cascade Lasers (ICLs) are semiconductor lasers emitting in the mid-wave infrared (MWIR 3-6 µm) and can operate as frequency combs (FCs). These demonstrations are based on double section cavities that can reduce dispersion and/or are adapted for radio-frequency operation. Here we show that ICLs FCs at long wavelengths, where the refractive index dispersion reduces, can be realized in a single long section cavity. We show FC generation for ICLs operating at λ ~ 4.2 µm, demonstrating narrow electrical beatnotes over a large current range. We also reconstruct the ultrafast temporal response through a modified SWIFT spectroscopy setup with two fast MWIR detectors, which shows a frequency modulated response in free-running operation. Further, we show that, through active modelocking, the ICL can be forced to generate short pulses on the order of 3 ps. This temporal response is in agreement with Maxwell Bloch simulations, highlighting that these devices possess long dynamics (~100ps) and potentially makes them appropriate for the generation of large peak powers in the MWIR.**

Interband Cascade Lasers (ICLs)[1] operate in the wavelength range of 3 µm to 6 µm, commonly known as the Mid-Wave Infrared (MWIR)[2]. These are of particular interest for gas spectroscopy applications, as many molecular species, including greenhouse gases, have strong absorption lines in the MWIR, as well as potential applications free-space optical communications, since the MWIR has a transparency window around 4 µm. The typical ICL output powers are on the order of a few to ten milliwatts with the low electrical power, enabling battery-powered and portable operation. With these complementary features compared to Quantum Cascade Lasers (QCLs)[3], ICLs are emerging as a reliable room temperature MWIR laser source[4].

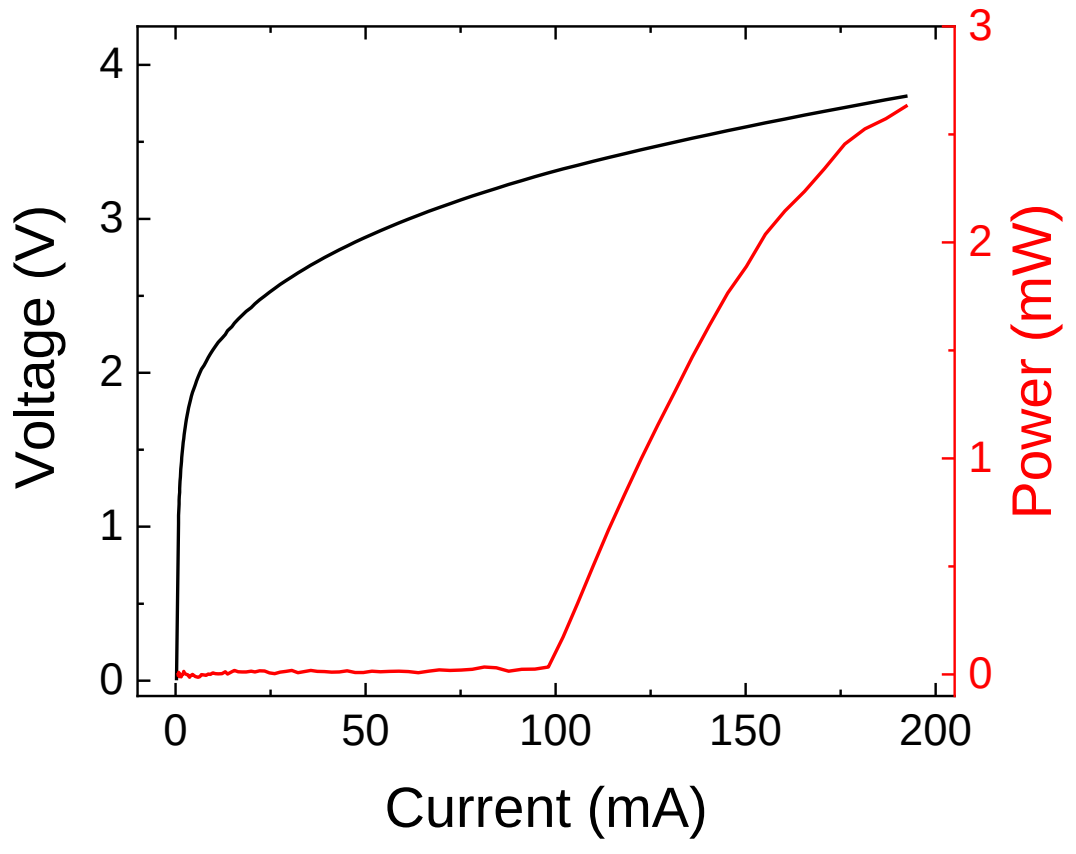

(a)

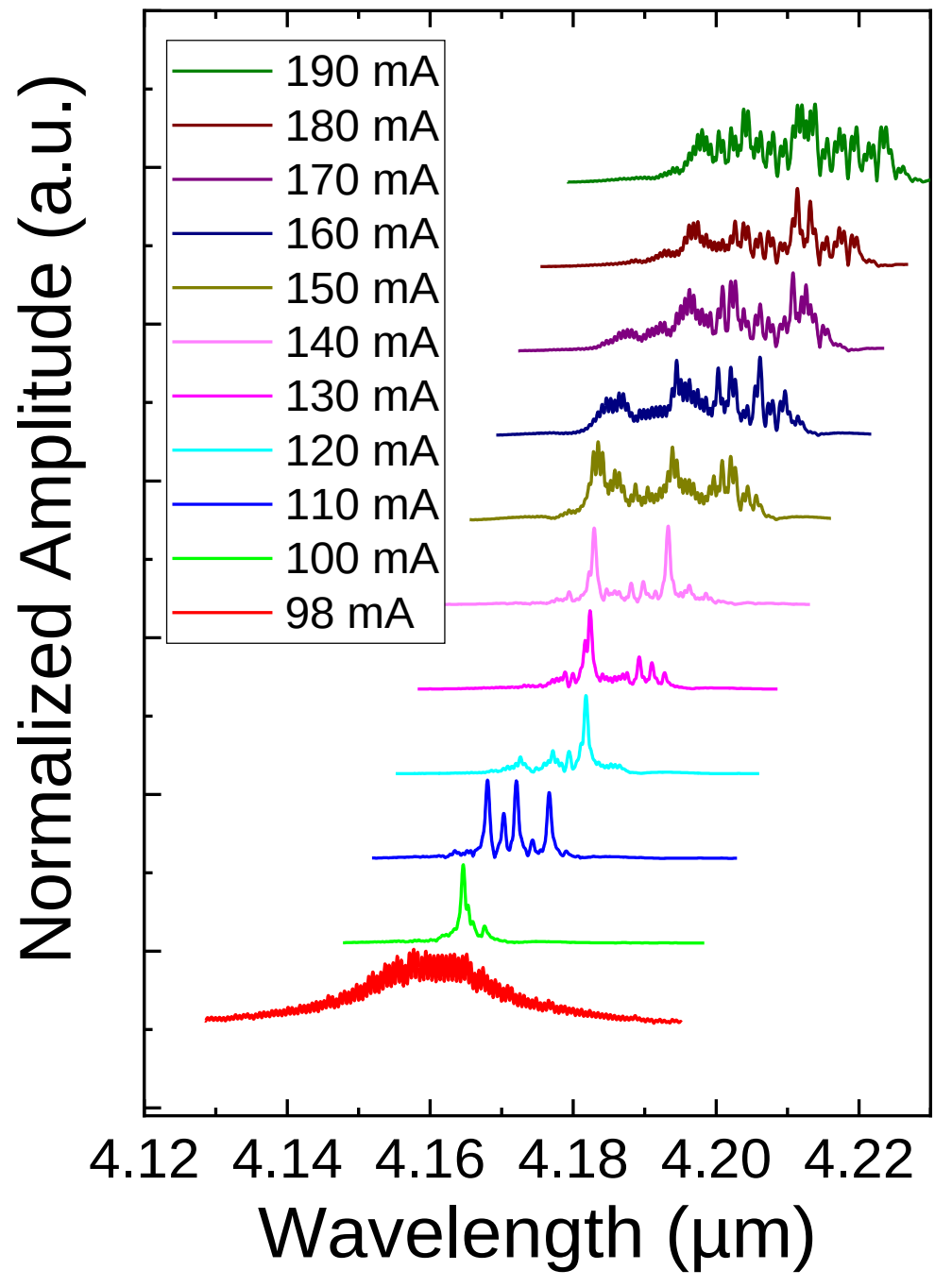

(b)

**Figure 1.** a) LIV characteristics of ICL operating at 4.2µm, 15°C in CW. The length of the ICL is 4mm with a ridge width of 10µm.b) Measured CW Spectrum of the ICL taken at 15°C from just below threshold to 190 mA. The spectra are normalised and offset for clarity

Furthermore, recent studies have shown that ICLs can operate in the frequency comb

(FC)[5–7] regime, offering further potential of ICLs for MWIR-based FC applications[8]. However, these works have emphasised the use of double section cavities for FC operation, optimized for RF injection/extraction, dispersion control or to control the lifetime in a section. Here we show that this is not entirely necessary and single section ICLs can be modelocked when the group velocity dispersion is reduced at long wavelengths. Indeed, at longer wavelengths, further from the bandgap resonance, the refractive index dispersion reduces and leads to low dispersion. This opens up the possibility of obtaining FC operation in ICLs without the implementation of dispersion compensation methods, simplifying the realisation of ICL FCs.

ICLs were realized for operation at $\lambda \sim 4.2$ μm, grown by molecular beam epitaxy on GaSb substrates[9]. The ICL structure consists of 7 active stages sandwiched between two 500 nm Te doped ($5 \times 10^{16}$ cm$^{-3}$) GaSb separate confinement layers (SCLs) along with top and bottom n-doped ($8 \times 10^{17}$ cm$^{-3}$) InAs/AlSb superlattice-based claddings of 2 μm and 3.5 μm thickness, respectively. The 'W' region (where the interband transition occurs) corresponds to quantum wells of InAs(1.9 nm)/$Ga_{0.65}InSb$ (3 nm)/InAs (1.6 nm) with hole and electron injectors similar to that described previously[10]. The structures were processed into 10 μm wide ridge Fabry-Perot lasers using standard photolithography. The narrow ridges were etched down to the bottom cladding by inductively-coupled plasma reactive-ion etching (ICP-RIE). Isolation and protection were then provided by hard baked AZ1518 photoresist. Contacts were made by depositing Ti/Au on top of the mesa and on the thinned substrate. Finally, the lasers were cleaved to create 4 mm long cavities without facet coating, and soldered episide-up with indium onto copper sub-mounts. The samples were wire bonded to a high speed coplanar waveguide, with the bonding at one end of the ICL, where the electrical beating is most susceptible to the RF injected signal due to its inherent spatio-temporal pattern[11,12]. These were placed on a Peltier cooler for temperature stability and tested at 15°C in CW. The typical Light-Current-Voltage curves (Figure 1a) show a threshold current of ~ 100 mA and a maximum output power of 2.4 mW at 193 mA

The normalized spectra taken with an FTIR are shown in Figure 1b from just below laser threshold (98mA) to 190 mA, showing a Stark shift from 4.16µm to 4.21µm with applied current. At currents greater than 140mA, the bandwidth increases significantly with a large number of Fabry-Perot modes. As mentioned above, at this wavelength the dispersion is usually low as illustrated in Figure 2a, which shows the refractive index and the group delay dispersion as a function of wavelength of InAs[13]. (InAs is chosen here for demonstration as the ICLs consists considerably of InAs). At $\lambda$ = 4.2 µm (ICLs that are studied here), the dispersion is considerably less than between 3.3 and 3.8 µm, where most studies of ICLs FCs have been performed. To demonstrate that the dispersion is indeed low at these wavelengths, the gain and dispersion were measured below laser threshold (94mA) (Figure 2b) using the interferogram and its satellite peak of the ICL electroluminescence with an FTIR. The satellite peak corresponds to the photon roundtrip of the ICL cavity[14]. The measured gain is centred at 4.16µm at this currents, with a dispersion of less than 1500 fs$^2$/mm in the typical bandwidth of operation of the ICL. This is considerably less than measured at shorter wavelengths[6] were most of the work on ICL FCs has been performed.

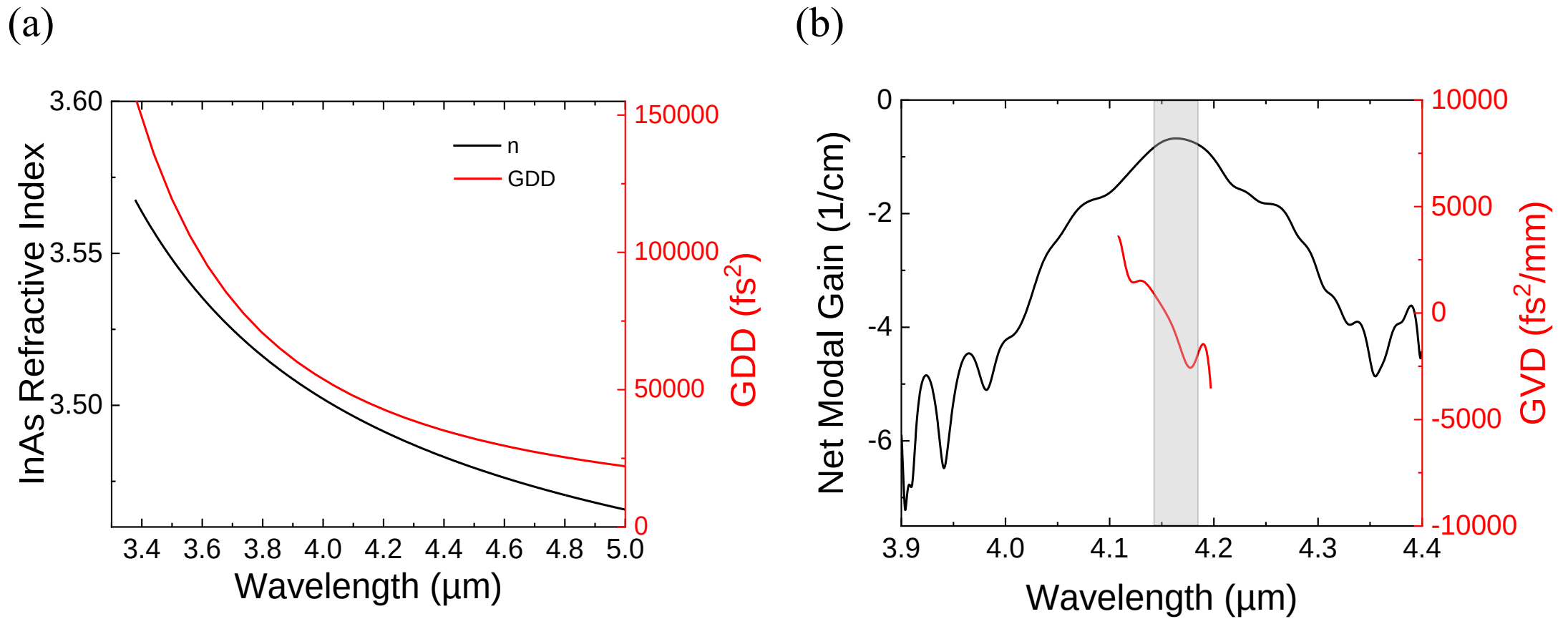

**Figure 2.** a) Refractive index and Group Delay Dispersion of InAs, showing that both reduce at longer wavelengths. b) Measured gain (log scale) and dispersion taken with a drive current of 94mA at 15°C, with the latter showing a low GVD

Before applying SWIFT spectroscopy to measure the temporal trace of the laser, the beatnote spectrum, measured electrically from the laser and extracted using an RF-compatible bias-tee, was analyzed as a function of current without applied RF modulation. Here, the modes of the ICL interfere to generate an RF signal at the ICL round-trip frequency $f_{RT}$ (around 10 GHz for a 4 mm cavity). The resulting map is shown in Figure 3a, demonstrating clearly the generation of a narrow beatnote from 135 mA to 190 mA that shifts from 9.875 GHz to 9.865 GHz. The narrowest beatnote was measured around 150 mA that demonstrated a FWHM of 40 kHz. This is the first indication that the ICL is indeed operating as a FC. The beatnote appears over almost over the entire current operating range and is a further indication of the low dispersion in this long wavelength ICL. The injection locking of this beatnote was then studied where a 17.5 dBm RF modulation is applied to the ICL that permits the ICL beatnote to be locked to a RF reference[15]. The beatnote map is shown in Figure 3b where the beatnote spectrum and intensity are plotted on varying the applied RF modulation between 9.8694 GHz and 9.8764 GHz. The acquisition of RF spectra is performed optically using fast Quantum Well Infrared Photodetectors (QWIPs). The ICL beatnote is observed close to 9.8716 GHz and is clearly 'pulled' towards the applied RF modulation as the latter's frequency is close to the ICL beatnote, and indicates that the ICL beatnote is locked to the external RF frequency[16,17]. Here the locking range – the frequency range of the ICL beatnote that is locked to external RF modulation – is 2.3 MHz. (This is relatively small owing to the fact that the ICL chip is not adapted for RF modulation and long bond wires are used, resulting in inefficient RF injection).

A new technique[18] has been implemented to identify whether there are different operating regimes during injection at different frequencies. We measured the amplitude and phase of the signal from QWIP with respect to the RF injection signal with a fast lock-in-amplifier (Zurich Intruments, HF2LI 50 MHz) using the same synthesizer for demodulation. RF signals are down-converted by a second RF signal generator to be compatible with the lock-in amplifier. Figure 3c shows that during injection, there is a 90° phase shift between the RT frequency and RT+1 MHz. This phase shift clearly indicates a change in the comb's operating regime, depending on whether the injection is in or out of phase with the RF beating[6,19,20]. These aspects will be further explored during the temporal characterization later in the paper.

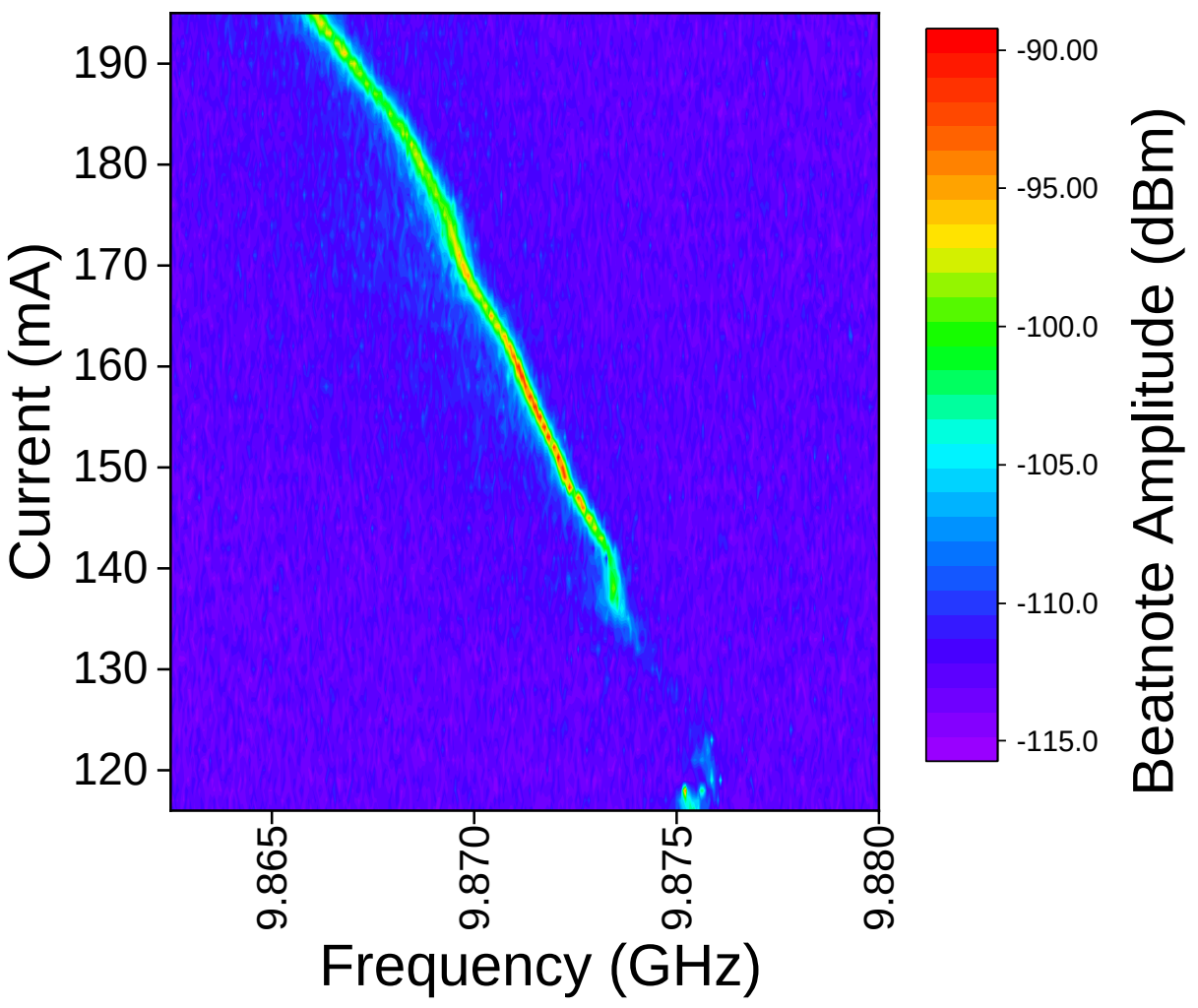

(a)

(b)

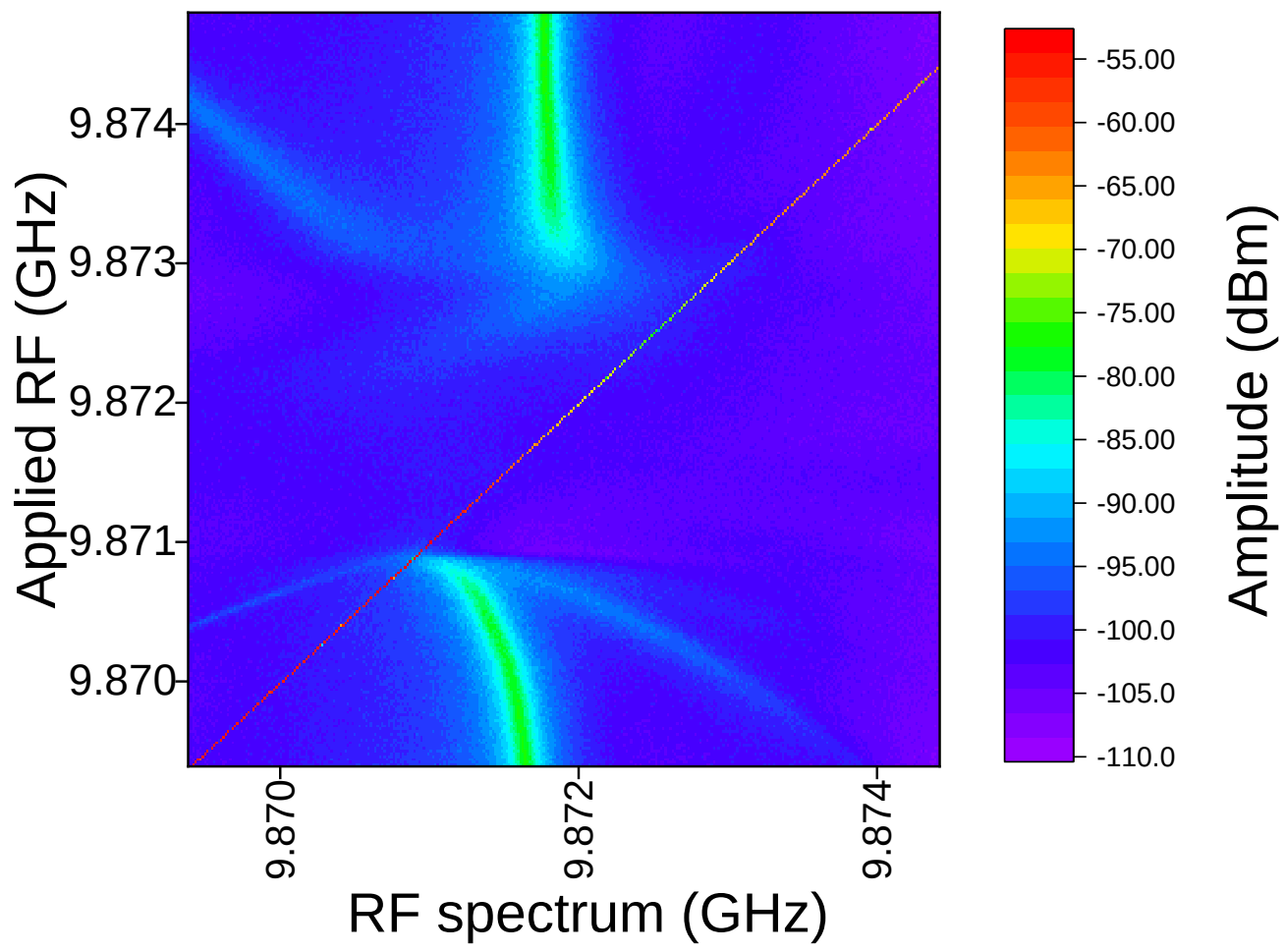

(c)

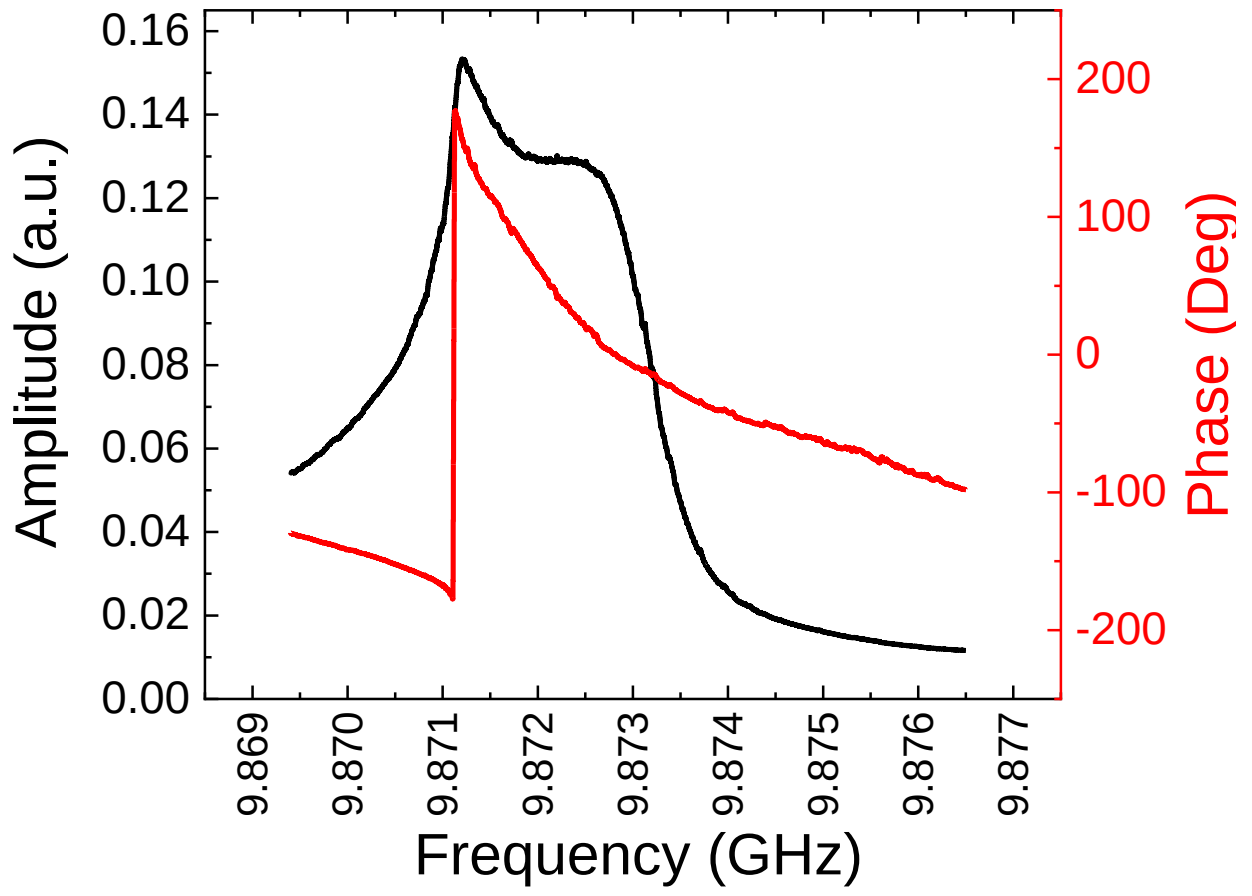

**Figure 3.** a) Measured beat-note measured electrically on the laser at 15°C in CW, showing a narrow beatnote over entire dynamic range of ICL. b) Injection locking of the ICL to a microwave reference. The ICL is operated at 150mA at 15°C with a beatnote around 9.8716 GHz, which can be locked to an external RF modulation over a locking range of ~ 2.3 MHz, This map is measured on QWIP with RBW 300 Hz. **c**) Measured amplitude and phase during RF injection. The amplitude corresponds to the diagonal of Figure 3b.

In the mode-locking region, there is a relatively rapid phase change. The maximum amplitude occurs very close to the RT frequency (at -0.3 MHz), which can be attributed to the electrical connections and the laser already operating at the natural frequency, equal to the beat note. However, the maximum injection efficiency does not necessarily imply that the optical modes are well in phase during injection at frequencies near RT. At higher frequencies, the amplitude curve reveals a second, less pronounced local maximum located at RT+1 MHz. This phase shift, combined with the presence of a secondary local maximum, indicates a change in the operating regime.

To demonstrate FC operation and reconstruct the temporal profile, a modified SWIFTS (Shifted Wave Interference Fourier Transform Spectroscopy)[21] system was developed, where $f_{RT}$ is measured and used to determine the amplitude and phase of the FC modes. This technique permits to demonstrate than the laser under test is operating as a FC. Although a coplanar waveguide is used for bonding, the ICL itself is not adapted for RF measurements and the extraction of the electrical beatnote directly from the ICL is not efficient. Instead we used two fast QWIPs with one to measure directly the beatnote and the second placed after the FTIR to measure the beatnote interferogram. A schematic of the setup is shown in Figure 4 where an RF oscillator, $f_{LO}$, is used to electrically stabilize the ICL. The ICL emission is then split into two parts, with the majority of the light sent through a FTIR spectrometer and beatnote detected by an adapted fast QWIP (2). This is then down-converted to ~ 25 MHz and sent to a fast lock-in amplifier. The small pick-off of the ICL emission is sent to a second QWIP (1) to measure the ICL beatnote and used as a reference for the lock-in amplifier after being down-converted ($f_{RT}$-$f_{LO}$). The QWIPs were designed to have a maximum spectral response at 4.2 µm (3.5 nm InGaAs quantum wells) and processed into mesa of dimensions of 80µm x 80 µm with an air-bridge contact for a fast response[22]. The 3 dB cut-off frequency was ~ 10 GHz. The QWIPs were operated at room temperature. (See supplementary material (SM) for further details).

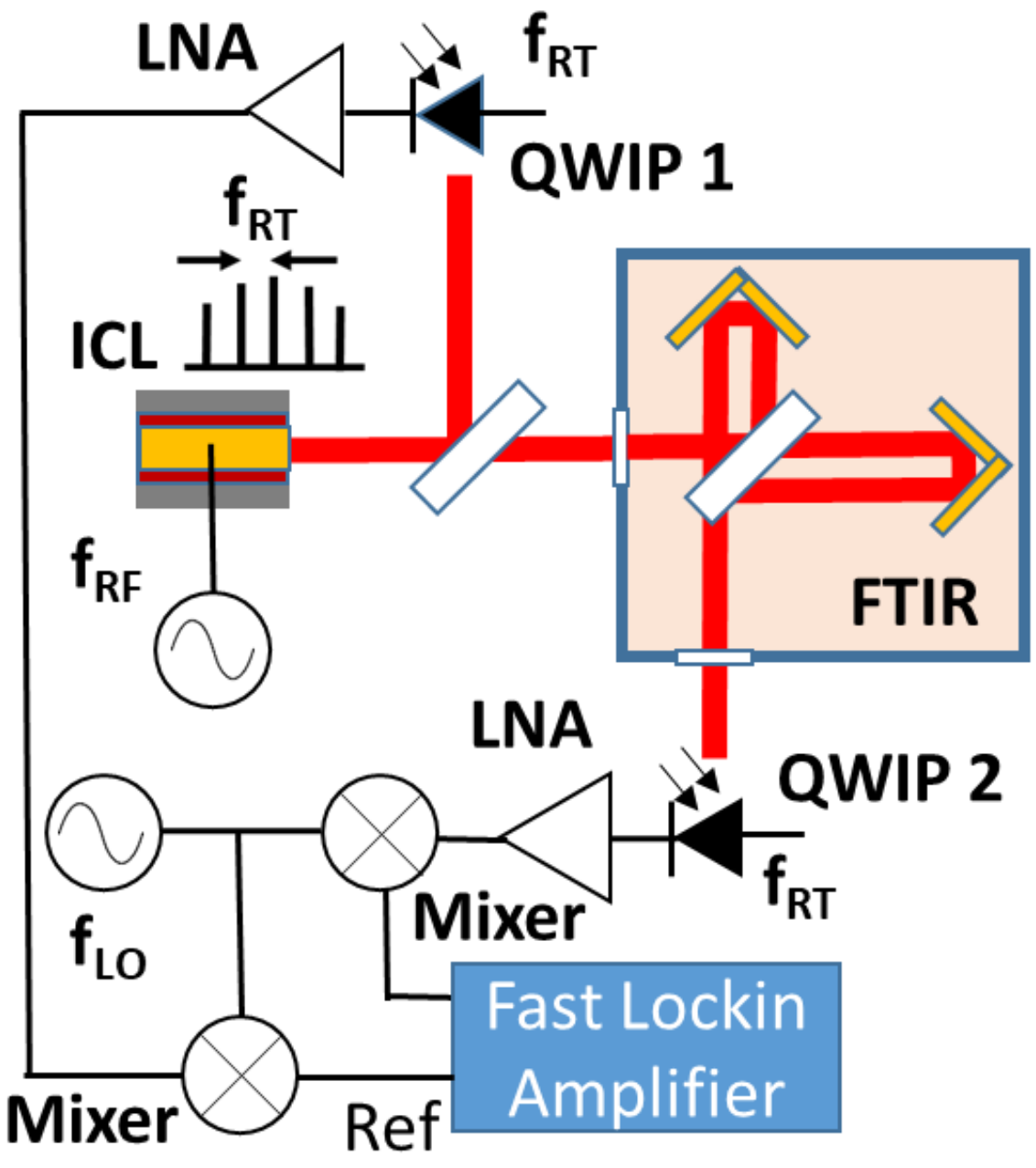

**Figure 4.** SWIFTS setup showing the use of two fast QWIPs to reconstruct the amplitude and phase of the ICL emission. QWIP 1 is used to provide a reference frequency for the lock-in amplifier while QWIP 2 is for the detection of the beatnote interferogram.

With narrow beatnotes and injection locking demonstrated, the temporal response was investigated using SWIFT spectroscopy in two cases of weak (-6 dBm) and strong (23 dBm) RF modulation, corresponding to free running and active modelocking operation, respectively. Figure 3 show the spectral amplitude and modal phase difference, the reconstructed temporal behavior of the intensity and the instantaneous frequency that are extracted from the SWIFTS quadrature interferograms. The intensity in Figure 3 (b,e,h) is normalized to the laser average power.

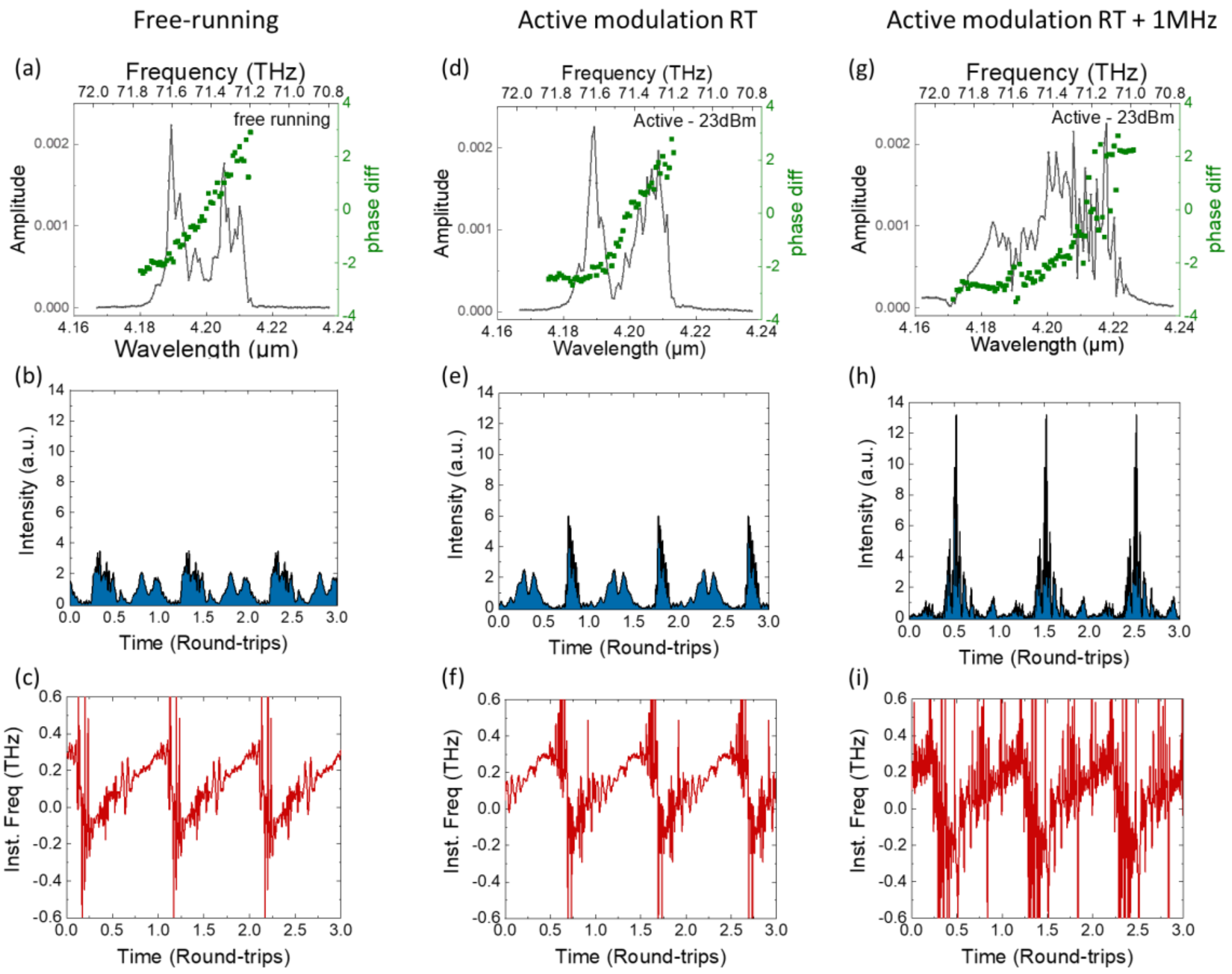

**Figure 3.** Reconstruction of temporal response of free running ICL (a, b, c), active modulated ICL at 23dBm at RT frequency (150mA) (d, e, f) and active modulated ICL with 1MHz offset from RT frequency (g, h, i). The top column shows the spectral amplitude and phase, the middle column the reconstructed intensity as a function of number of round trips, and the bottom column is the instantaneous frequency. One round trip corresponds to 101 ps (9.8716 GHz for a 4mm long cavity).

Regarding the free running case, the acquired spectrum shown Figure 3a has the same form as the spectrum in Figure 2a but with the added information of the phase that shows a variation over $2\pi$. This is a clear indication of chirp as observed in previous reports of ICLs that operate at shorter wavelengths. The reconstructed time profile and instantaneous frequency from the amplitude and phase are shown in Figure 3 b and c with the former showing the intensity distributed over one cavity round trip and the latter clearly showing a frequency chirp. This shows that the ICL is frequency modulated in the free running case. At higher RF modulation (23 dBm), we clearly start to see a change in behavior of the ICL as it transits to amplitude modulation behavior. The intensity of the modes changes in the spectrum with a strong effect on the phase (Figure 3d) with the first lobe showing a flat phase and the second showing a chirped response. This is then observed in the time and instantaneous frequency (Figure 3 e and f), where we see a two lobes with the first corresponding to the flat phase and the second corresponding to the chirp seen in Figure 3d. This observation shows that the two lobes are anticorrelated; indeed, if we look at the spectral lobe with a flat phase, it appears in the time domain with a high amplitude typical of AM characteristics, while the other lobe shows a response typical of FM. The phase difference between the two lobes is approximately π. Such observations of anti-correlation phenomena have been previously reported in the case of QCLs[23].

Finally, an extreme of an amplitude modulated FC is observed in the case of the ICL operating at slightly higher currents and a RF modulation frequency detuned by only 1MHz from the round trip frequency. This introduces in-phase synchronization[6,19,20] between modes and allows a transition to amplitude modulation behavior, where detuning has shown to help in the generation of pulses in QCLs as it forces the laser to operate away from its natural frequency modulation state[6,20,24]. As can be observed (Figure 3 g) the spectrum is broader and the majority of modal phases are flat. This clearly translates into a pulse behavior with pulses of duration of 3.2 ps and with no or little chirp response (Figure 3 h and i). The peak power represents an enhancement of more than 10 with respect to the average power. Note that there is a significant atmospheric absorption from $CO_2$ at wavelengths ≳ 4.2 μm that could affect the form of the pulses. (Nonetheless the absorption effects are reduced compared to the classical SWIFTS scheme with one QWIP, as presented in the first version of the SWIFTS setup[25]).

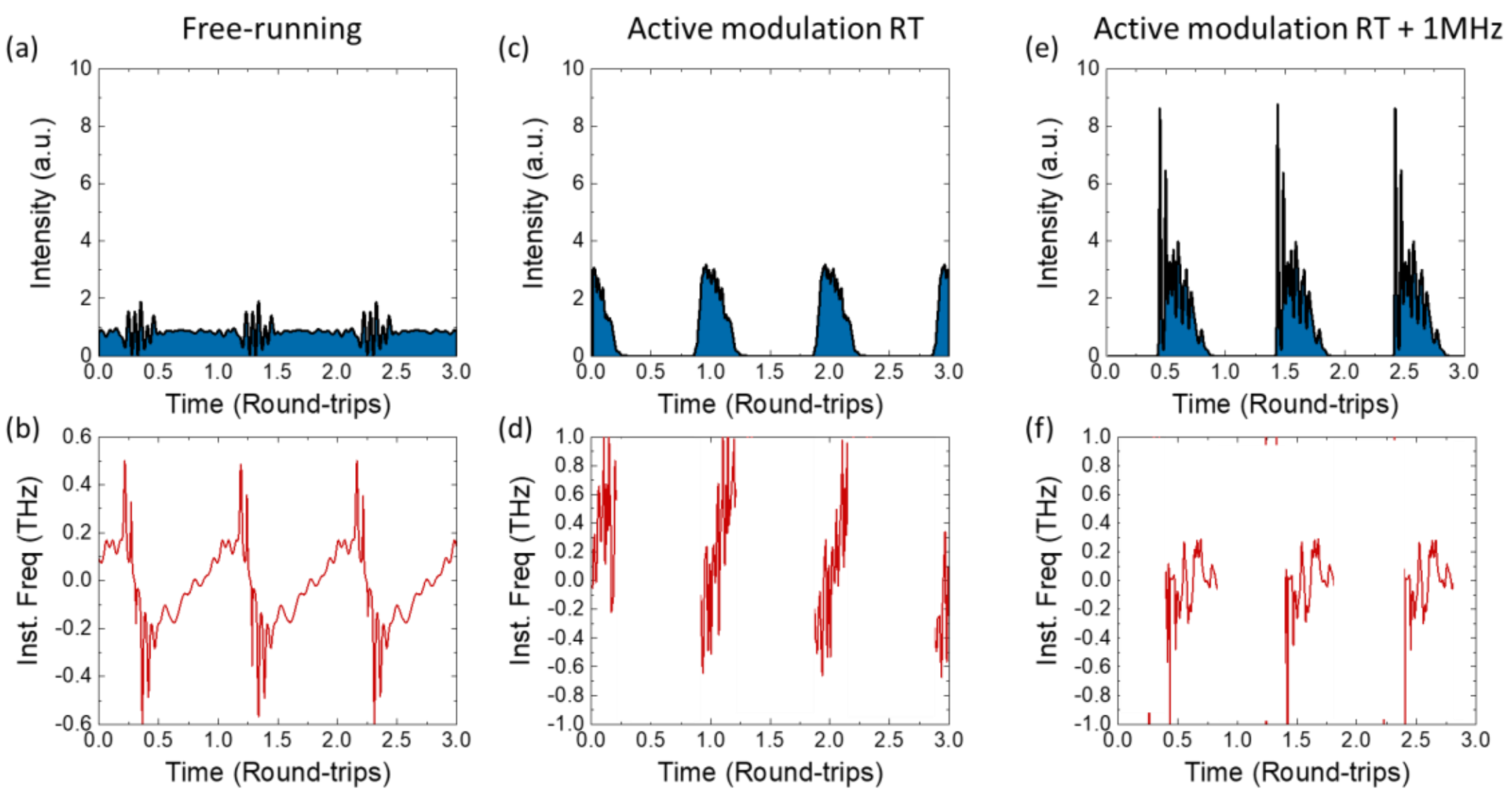

**Figure 4.** Simulated time profiles of the ICL in the case of free running, active modulated at the RT frequency and active modulated at the RT frequency with a small offset of 1 MHz. The top row is the intensity of the emission, and the bottom is the instantaneous frequency as a function of number of round trips.

These results were then compared to Maxwell-Bloch simulations which describe the full electric-field dynamics in the spatially extended laser cavity. We employ a travelling-wave model based on delay-algebraic equations, an approach that has been successfully applied in the past[26,27] and is described in further detail in the SM. Spatial hole burning (SHB), the mechanism responsible for the creation of the FM behavior of semiconductor laser emission[28], is taken into account through the spatial inhomogeneity in the available gain due to the spatial beating between the counter-propagating waves in the laser cavity. Regarding SHB, this is expected to efficient as the carrier diffusion time is expected to be longer than the lifetime of the laser transition, as is the case for QCLs. To numerically investigate the impact of RF injection onto the laser dynamics we only modulate a gain section close to one facet of the cavity as in the experiment. This is where the electrical beating is most susceptible to the injected signal due to its inherent spatio-temporal pattern[11,12]. The results are shown in Figure 4 that shows the intensity (top) and instantaneous frequency (bottom column) as a function of time. Similar tendencies are observed a frequency modulated response for free running and more pulse like behavior for the RF modulated. The simulations also can replicate the results observed in Figure 3 where clearer pulses are observed with a slight detuning of the modulation frequency. Interestingly the lifetime plays a critical role in the active modelocking response. Here, the closest fit to the data in the case of modulation at the round-trip is with a lifetime of 100 ps and in agreement with the slow dynamics that have been measured of the laser transition itself[29]. Although fast dynamics have been measured in the carrier injection, this does not appear a factor in the current simulations for active modulation and suggests that energy can be stored within the cavity an open up prospects for passive modelocking.

In this work we have shown modelocked operation and pulse generation of a single section ICL, highlighting FC operation without the need of double section, RF contacts or more complex structures. This is achieved by realizing FC ICLs at longer wavelengths than previously investigated where the dispersion of the material is inherently less. A modified SWIFT setup was realized to measure the ultrafast characteristics. The temporal characterisation of the ICLs show a frequency chirped response in free running operation and a pulsed behavior with a relatively low RF injected power $< 200$ mW. These results were compared to Maxwell-Bloch simulations in the free running and actively modelocked cases that showed a good agreement and the important role of the gain lifetime for the formation of pulses. As the dispersion appears to play an important role, this can be further reduced by employing large gain ICLs or going to longer wavelengths[30]. Further a simple and existing coplanar type geometry could be realised using side contacts as already demonstrated in ICLs on different type of substrates[9]. This would be of interest to improve the injection efficiency of the RF modulation for the formation of shorter pulses with higher peak powers. This work therefore will be particularly important in the application of ICLs as a compact FC source for the MWIR region.

The authors acknowledge funding from the French National Research Agency (ANR-19-CE24-0023 - “ACTIVE-COMB” and ANR-11-EQPX-0016) and was partly supported by the French RENATECH network coordinated by the CNRS.